\documentclass[11pt]{article}
\usepackage{jheppub}
\usepackage{epsfig}
\usepackage{amsmath,amssymb}
\usepackage{amsfonts}
\usepackage{mathrsfs}
\usepackage{relsize}
\RequirePackage{xspace}
\usepackage{graphicx}
\usepackage{slashed}
\usepackage{bm}

\title{ Study on Toponium: Spectrum and Associated Processes }

\author[a]{ Sheng-Juan Jiang }
\author[b]{,Bai-Qing Li  }
\author[a]{,Guang-Zhi Xu }
\author[c]{,Kui-Yong Liu  }

\affiliation[a]{School of Physics, Liaoning University, Shenyang 110036
, China}

\affiliation[b]{School of Science, Huzhou University, Huzhou 313000
, China}

\affiliation[c]{School of Physics, Liaoning Normal University, Dalian 116029
, China}

\emailAdd{xuguangzhi@lnu.edu.cn}
\emailAdd{liukuiyong@lnu.edu.cn}

\date{\today}

\abstract{
In this paper, we calculate the toponium spectrum in the potential model with the screened effects. Coulombic part is dominant for toponium, and the coefficient of the Coulomb potential is chosen from lattice QCD calculations at an infinite quark mass in accord with the ultrasoft scale choice. We also derive the relative E1 transitions, decays, and production.
}

\begin{document}
\maketitle
\flushbottom

\section{Introduction}

The top quark is the heaviest one among all elementary particles and has an extremely short lifespan, making it difficult to form a bound state containing a top quark. On the other hand, since the discovery of the first heavy quark meson $J/\psi$ in 1974, heavy quarkonium (including charmonium and bottomnium), as a relatively simple two-body system and an excellent platform for studying perturbative and non-perturbative chromodynamics, has inspired numerous experimental and theoretical studies over the past five decades. 

Recently, the CMS and ATLAS collaborations reported an excess in the data for invariant mass of $t\bar{t}$ near the threshold, which is preferring a pseudoscalar over a scalar hypothesis\cite{CMS:2024ynj,ATLAS:2023fsd,Jeppe:2024uki}. Meanwhile, several theoretical works predicted peaks in the
$t\bar{t}$ invariant-mass distribution below or near the thresold\cite{Hagiwara:2008df,Kiyo:2008bv,Fuks:2021xje,Sumino:2010bv}. And the experimental data excess supports QCD toponium rather than some contact interactions beyond the Standard Model\cite{Llanes-Estrada:2024phk}.
It is certain that if toponium exists, it will have the smallest scale and the simplest structure due to the large mass of the top quark\cite{Fu:2024bki}.
Therefore,  compared to the long-range interaction, the short-range one is of greater importance. The mass spectra of toponium can be simply analogized to the energy levels of the hydrogen atom or positronium.
Compared with it, the higher-mass states above the open-flavor threshold for charmonia and bottomonia, which are not well understood, force ones to modify the long-range part in the potential model \cite{Brambilla:2019esw,Chen:2022asf,Chen:2016spr}, such as introducing a screened potential instead of the linear one\cite{Li:2009zu,Li:2009nr}. 
Toponium spectra were revisited based on the instantaneous Bethe-Salpeter equation\cite{Wang:2024hzd} and non-relativistic potential model \cite{Akbar:2024brg} in which Coulomb plus a linear confinement potential is adopted and the parameters fit to the lowest mass from the data. In this paper, we also investigate the toponium mass spectra. In the calculations, the coefficient of the Coulomb potential is derived from lattice QCD, and the effect of the screening potential is also incorporated. Furthermore, the results of some decay and production processes are also presented.

The rest of this paper is structured as follows. In Section 2, we derive the mass spectra. In Section 3, the electric dipole transitions are calculated. In Section 4, we compute the decay widths of leptonic decays, multi-photon decays, and multi-gluon decays. In Section 5, the cross sections of $\eta_t$ production in proton-proton colliders are determined. Finally, a brief summary is provided.

\section{Spectra}

\label{sec:Spectra}

We employ the screened potential form to investigate the toponium spectra. The potential is given by
\begin{equation}\label{eq:potential}
V_{\text{scr}}(r) = V_V(r) + V_S(r),
\end{equation}
where
\begin{equation}
V_V(r) = -\frac{\lambda}{r},
\end{equation}
and
\begin{equation}
V_S(r) = \sigma \left( \frac{1 - e^{-\mu r}}{\mu} \right).
\end{equation}
Here, the coefficient $\lambda$ of the Coulomb potential is derived from lattice QCD calculations at an infinite quark mass \cite{Kawanai:2013aca,Fu:2024bki}, with $\lambda = 0.2875$. This value is compared with soft scale choice ($0.20343$ from Ref.~\cite{Kiyo:2008bv}, $0.2128$ from Ref.~\cite{Akbar:2024brg}) and in accord with the ultrasoft scale (eg. $0.3138$ from Ref.~\cite{Llanes-Estrada:2024phk}).
The string tension coefficient is selected as $\sigma = 0.21~\text{GeV}^2$.
The screening factor $\mu$ causes the long-range part of $V_S(r)$ to decay exponentially. It represents the energy scale (or the length scale $1/\mu$) associated with the creation of light quark pairs. By fitting the spectrum, $\mu = 0.0979~\text{GeV}$ for charmonium and $\mu = 0.056~\text{GeV}$ for bottomonium \cite{Li:2009nr,Li:2009zu}. It indicates that, in comparison with charmonium, the screening effect attains significance at a relatively larger distance for bottomonium\cite{Li:2009nr}. 
In the case of toponium, due to the exceedingly small distance between $t\bar{t}$ pair, light quark pairs are likely to be created at even more substantial distances. Consequently, a relatively smaller value of \(\mu = 0.03~\text{GeV}\) is designated in our calculations.

The spin-dependent interactions, which consist of three terms (denoted as $H_{SS},H_{LS},H_{T}$ for spin-spin, spin-orbit, and tensor items, respectively) are  incorporated to acquire the hyperfine and fine energy level structure.
\begin{equation}\label{eq:ingerterm}
  H_{SS}+H_{LS}+H_{T}=\frac{8\pi\lambda}{3m_t^2}\widetilde{\delta}(r)\vec{S}_t\cdot\vec{S}_{\bar{t}}
  +\frac{1}{2m_{t}^{2}r}(3V_{V}'(r)-V_{S}'(r))\vec{L}\cdot\vec{S}+\frac{1}{12m_{t}^{2}}\left(\frac{1}{r}V_{V}'(r) - V_{V}''(r)\right)T,
\end{equation}
where $\widetilde{\delta}(r)$ is usually taken as Dirac $\delta(\vec{r})$ in nonrelativistic potential models. 

In a $|J,L,S\rangle$ basis, the matrix elements of spin-spin, spin-orbit operator is diagonal presented as follows,

$$\langle\vec{S}_t\cdot\vec{S}_{\bar{t}}\rangle=\frac{1}{2}S^2-\frac{3}{4} = \frac{1}{2}S(S+1)-\frac{3}{4} = 
\begin{cases} 
-\frac{3}{4}, & S=0 \\ 
\frac{1}{4}, & S=1
\end{cases} $$

$$\langle\vec{L}\cdot\vec{S}\rangle=\left[J(J+1)-L(L+1)-S(S+1)\right]/2$$
For the tensor operator, the matrix element possess non-zero values only for the $L>0$ spin-triplet states, with the common form,
\begin{equation} 
\langle ^3L_J | T | ^3L_J \rangle = 
\begin{cases} 
-\frac{L}{6(2L+3)}, & J = L + 1 \\ 
\frac{1}{6}, & J = L \\
-\frac{(L+1)}{6(2L-1)}, & J = L - 1. 
\end{cases} 
\end{equation}

For toponium, the Coulombic part in the potential is the dominant and the corrected mass or other parameters from all the other parts can be considered as perturbations. With the potential form (Eq.~\ref{eq:potential}), 
the averaged radii for $S,P,D$-wave states 
are given in Table.\ref{tab:r2}. It can be verified that the proportional relationship between these radii is nearly the same as that of the hydrogen atom or positronium. The Bohr radius $r_B$ can be derived by virtue of its relationship with the average radius of the $1S$ state, ie. $r_B=\frac{2}{3}\langle r\rangle_{1S}$ (also  related to the coefficient $\lambda$ in the Columbic potential, $r_B=\frac{2}{m_t\lambda}$ 
). Compared with charmonium and bottomnium, the values are more than one order of magnitude smaller. 
This results verify that toponium is an extremely small bound state on the spatial scale.

{
\begin{table}[!hbp]
 \begin{center}
\caption{\label{tab:r2} The averaged radii for different orbital quantum numbers in units of $10^{-2}~\textrm{fm}$.}
\begin{tabular}{c|cccccc|ccc|ccc}
\hline
\hline
&$1S$&$2S$&$3S$&$4S$&$5S$&$6S$&$1P$&$2P$&$3P$&$1D$&$2D$&$3D$ \\
$\langle r^2\rangle^{1/2}$&$1.39$&$5.07$&$10.51$&$16.58$&$22.54$&$28.17$&$4.27$&$9.74$&$15.83$&$8.11$&$14.37$&$20.42$ \\
\hline
$\langle r\rangle$&$1.20$&$4.70$&$9.86$&$15.58$&$21.14$&$26.35$&$3.91$&$9.07$&$14.80$&$7.63$&$13.42$&$19.04$ \\
\hline
\end{tabular}
\end{center}
\end{table}}

Due to the heavy top mass, 
the mass splitting between the singlet and triplet states is negligible except $1S,2S$ orbital, which is the hyperfine structure from spin-spin contact interaction. 
Here, the effects of the spin-spin contact interaction 
are considered as perturbations to the binding energy or the mass \cite{A:2023bxv}.
Integrating out the $\widetilde{\delta}(r)$ function of the first term of the interaction in Eq. \ref{eq:ingerterm}, 
an estimation from the mass splitting for $1S$ states is written as follows \cite{Frank:1985pn},
\begin{equation}\label{eq:massspl}
  \Delta M_{SS}^{1S}=\frac{8\pi\lambda}{3m_t^2}\frac{|R_{1S}(0)|^2}{4\pi}\simeq\frac{8\pi\lambda}{3m_t^2}\frac{1}{\pi r_B^3}A,~~A=\begin{cases} 
-\frac{3}{4}, & S=0 \\ 
\frac{1}{4}, & S=1
\end{cases}.
\end{equation}
where $S$-wave radial functions at the origin satisfies approximate relationship with Bohr radius, $|R_{1S}(0)|^2\simeq\frac{4}{r_B^3}$.
The mass splitting of the $2S$ state is one-eighth of the value of the $1S$ state.
We notice that the values of the mass splitting are significantly larger than those for charonium and bottomnium. 
    
\begin{figure}[htbp]
  \centering
  \includegraphics[width=1\textwidth]{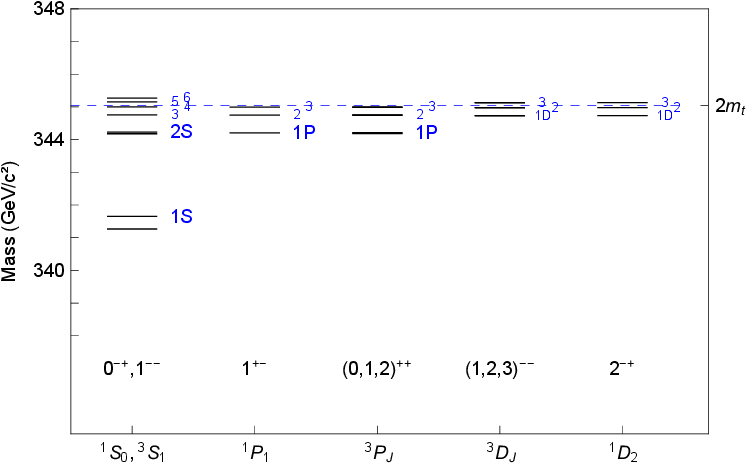}
  \caption{ Theoretical mass spectrum for toponium states.\label{fig:topniumfamily}}
\end{figure}

In the calculations, the top quark mass is taken from the current most precise measurement \cite{ATLAS:2024dxp}, $m_t = 172.52~\textrm{GeV}$.
The final mass spectra is shown in Table.~\ref{tab:mass} and also illustrated as Fig.~\ref{fig:topniumfamily}.
The binding energy of $nS$ states for different potential form is shown in Table.~\ref{tab:be}. The screened effects depress the mass of the higher excited state slightly in comparison with Cornell potential calculations.

{
\begin{table}[!hbp]
 \begin{center}
\caption{\label{tab:mass} Theoretical mass spectrum for toponium states in units of \textrm{GeV}.}
\scalebox{0.9}{
\begin{tabular}{c|c|c|c|c|c}
\hline
\hline
$1S(1^3S_1,1^1S_0)$&$2S(2^3S_1,2^1S_0)$&$3S(3^3S_1,3^1S_0)$&$4S(4^3S_1,4^1S_0)$&$5S(5^3S_1,5^1S_0)$&$6S(6^3S_1,6^1S_0)$ \\
$341.650,341.267$&$344.227,344.179$&$344.759$&$345.000$&$345.153$&$345.269$ \\
\end{tabular}
}
\scalebox{0.8}{
\begin{tabular}{cccc|cccc|cccc}
\hline
$1^1P_1$&$1^3P_0$&$1^3P_1$&$1^3P_2$&$2^1P_1$&$2^3P_0$&$2^3P_1$&$2^3P_2$&$3^1P_1$&$3^3P_0$&$3^3P_1$&$3^3P_2$ \\
$344.206$&$344.181$&$344.200$&$344.215$   &$344.751$&$344.742$&$344.749$&$344.754$    &$344.992$&$344.987$&$344.991$&$344.994$ \\
\hline
$1^3D_1$&$1^3D_2$&$1^3D_3$&$1^1D_2$&$2^3D_1$&$2^3D_2$&$2^3D_3$&$2^1D_2$&$3^3D_1$&$3^3D_2$&$3^3D_3$&$3^1D_2$ \\
$344.733$&$344.735$&$344.736$&$344.735$   &$344.975$&$344.976$&$344.977$&$344.976$   &$345.129$&$345.130$&$345.131$&$345.130$ \\
\hline
\end{tabular}
}
\end{center}
\end{table}}

{
\begin{table}[!hbp]
 \begin{center}
\caption{\label{tab:be} The binding energy (in units of \textrm{GeV}) of $nS$ states for different potential form.}
\begin{tabular}{c|cccccc}
\hline
potential&$1S$&$2S$&$3S$&$4S$&$5S$&$6S$\\
\hline
Coulomb&$-3.499$&$-0.876$&$-0.389$&$-0.219$&$-0.140$&$-0.097$ \\
Cornell&$-3.486$&$-0.825$&$-0.280$&$-0.038$&$0.117$&$0.235$ \\
screened&$-3.486$&$-0.825$&$-0.281$&$-0.040$&$0.113$&$0.229$ \\
\hline
\end{tabular}
\end{center}
\end{table}}

\section{Electric dipole (E1) transitions}

A non-relativistic formula for E1 transitions from the Ref.\cite{Kwong:1988ae} is written as follows , 
\begin{equation}\label{eq:e1tran}
  \Gamma(n{}^{2S+1}L_J\rightarrow n^\prime{}^{2S^\prime+1}L^\prime_{J^\prime}+\gamma) = \frac{4}{3}C_{fi}\delta_{SS^\prime}e_t^2|\langle f|r|i\rangle|^2E_\gamma^3
\end{equation}
where $E_\gamma$ denote the energy of emitted photon, and the spatial matrix element $\langle f|r|i\rangle$ is obtained by integrating the radial wave functions of the initial state and the final state, 

\begin{equation}\label{eq:spatialmaxele}
  \langle f|r|i\rangle = \int_{0}^{\infty}R_f(r)R_i(r)r^3dr.
\end{equation}

The angular matrix element $C_{fi}$ is 
\begin{equation}\label{eq:angmaxele}
  C_{fi} = \max(L,L^\prime)(2J^\prime+1)\left\{\begin{matrix}
    L^\prime & J^\prime & S \\
    J & L & 1 
  \end{matrix}\right\}^2
\end{equation}

Our results are listed in Table.\ref{tab:E1decay}. The widths for the channel $1P\to 1S\gamma$ are more than $60/90~\textrm{keV}$ in comparison with $20~\textrm{keV}$ in Ref.~\cite{Akbar:2024brg} mainly because of the mass difference for $1S$ states. 

{
\begin{table}[!hbp]
 \begin{center}
\caption{\label{tab:E1decay} E1 transitions rates of toponium states in units of $\textrm{keV}$.}
\begin{tabular}{ccccc}
\hline
\hline
states&Initial state&Final state&$E_\gamma(\textrm{GeV})$&$\Gamma(\textrm{keV})$ \\
$1P\rightarrow 1S$&$1^3P_0$&$1^3S_1$&$2.537$&$67$ \\
                &$1^3P_1$&$1^3S_1$&$2.556$&$68$ \\
                &$1^3P_2$&$1^3S_1$&$2.571$&$69$ \\
                &$1^1P_1$&$1^1S_0$&$2.882$&$98$ \\
$2P\rightarrow 1S$&$2^3P_0$&$1^3S_1$&$3.094$&$22$ \\
                &$2^3P_1$&$1^3S_1$&$3.100$&$22$ \\
                &$2^3P_2$&$1^3S_1$&$3.105$&$22$ \\
                &$2^1P_1$&$1^1S_0$&$3.422$&$29$ \\                
$2P\rightarrow 2S$&$2^3P_0$&$2^3S_1$&$0.516$&$3$ \\
                &$2^3P_1$&$2^3S_1$&$0.523$&$3$ \\
                &$2^3P_2$&$2^3S_1$&$0.528$&$3$ \\
                &$2^1P_1$&$2^1S_0$&$0.568$&$4$ \\
$2S\rightarrow 1P$&$2^3S_1$&$1^3P_0$&$0.044$&$2\times10^{-3}$ \\
                &$2^3S_1$&$1^3P_1$&$0.026$&$1\times10^{-3}$ \\
                &$2^3S_1$&$1^3P_2$&$0.011$&$1\times10^{-4}$ \\
                &$2^1S_0$&$1^1P_1$&$0$&$0$ \\
$3S\rightarrow 1P$&$3^3S_1$&$1^3P_0$&$0.582$&$0.1$ \\
                &$3^3S_1$&$1^3P_1$&$0.563$&$0.3$ \\
                &$3^3S_1$&$1^3P_2$&$0.548$&$0.5$ \\
                &$3^1S_0$&$1^1P_1$&$0.542$&$3$ \\ 
$1D\rightarrow 1P$&$1^3D_1$&$1^3P_0$&$0.552$&$6$ \\
                 &         &$1^3P_1$&$0.533$&$4$ \\
                 &         &$1^3P_2$&$0.528$&$0.3$ \\
                &$1^3D_2$&$1^3P_1$&$0.535$&$7$ \\
                 &         &$1^3P_2$&$0.520$&$2$ \\ 
                 &$1^3D_3$&$1^3P_2$&$0.521$&$9$ \\
                &$1^1D_2$&$1^1P_1$&$0.529$&$14$ \\                             
\hline
\end{tabular}
\end{center}
\end{table}}

\section{Decay Processes}

Subsequently, within the color singlet model, we present the widths of some decay processes and the cross sections of production processes of toponium. We can obtain the analytical calculation results of short distance ratios from previous works. The probability of $Q\bar{Q}$ pair hadronization into mesons is characterized by non-perturbative long distance matrix elements or radial wave functions. From the calculation of the above potential, we obtain the values of $S$-wave radial functions, first derivative of $P$-wave radial  functions and second derivative of $D$-wave radial functions at the origin as follows,
\begin{eqnarray}
  |R_{1S}(0)|^2 &=& 1.14784\times10^{-2}m_t^3=5.89387\times10^4~\rm{GeV^3} \nonumber\\
  |R_{2S}(0)|^2 &=& 1.51794\times10^{-3}m_t^3=0.77942\times10^4~\rm{GeV^3} \nonumber\\
  |R_{3S}(0)|^2 &=& 5.34588\times10^{-4}m_t^3=0.27450\times10^4~\rm{GeV^3} \nonumber\\
  |R_{4S}(0)|^2 &=& 3.00369\times10^{-4}m_t^3=0.15423\times10^4~\rm{GeV^3} \nonumber\\
  |R_{5S}(0)|^2 &=& 2.14083\times10^{-4}m_t^3=0.10992\times10^4~\rm{GeV^3} \nonumber\\
  |R_{6S}(0)|^2 &=& 1.71957\times10^{-4}m_t^3=0.08830\times10^4~\rm{GeV^3} \nonumber\\
  |R_{1P}^\prime(0)|^2 &=& 2.62747\times10^{-6}m_t^5=4.01545\times10^5~\rm{GeV^5} \nonumber\\
  |R_{2P}^\prime(0)|^2 &=& 1.11383\times10^{-6}m_t^5=1.70222\times10^5~\rm{GeV^5} \nonumber\\
  |R_{3P}^\prime(0)|^2 &=& 0.67270\times10^{-6}m_t^5=1.02806\times10^5~\rm{GeV^5} \nonumber\\
  |R_{1D}^{\prime\prime}(0)|^2 &=& 5.61841\times10^{-10}m_t^7=2.55558\times10^6~\rm{GeV^7}\nonumber\\
  |R_{2D}^{\prime\prime}(0)|^2 &=& 4.90184\times10^{-10}m_t^7=2.22964\times10^6~\rm{GeV^7} \nonumber\\
  |R_{3D}^{\prime\prime}(0)|^2 &=& 4.36460\times10^{-10}m_t^7=1.98527\times10^6~\rm{GeV^7}
\end{eqnarray}\label{eq:wavefuns}
Remarkably, the third-order corrections to the $S$ wave suggest important effects on the $S$-wave function \cite{Beneke:2007gj,Beneke:2024sfa}.

The value of the fine-structure constant $\alpha$ in electromagnetic processes is taken as $1/137$. For the strong running coupling constant $\alpha_s$, which is dependent on the scale $\mu$, a two-loop expression is chosen,
\begin{equation}\label{eq:alphas2loop}
  \frac{\alpha_s(\mu)}{4\pi} = \frac{1}{\beta_0 L} - \frac{\beta_1 \ln L}{\beta_0^3 L^2},
\end{equation}
where
$L=\ln(\mu^2/\Lambda_{QCD}^2)$, $\beta_0=(11C_A-4T_f n_f)/3$, $\beta_1=34C_A^2/3-4(5C_A/3+C_F)T_f n_f$ with $\Lambda_{QCD}=220~\textrm{MeV}$, and $n_f=5$ for the production and decay of toponium states. Therefore, we obtain $\alpha_s(2m_t)=0.098$.

\subsection{Leptonic decays}

The formulas of di-electronic decay width for the vector meson ($n^3S_1,n^3D_1$ states) are well known with the QCD
radiative corrections \cite{VanRoyen:1967nq,Barbieri:1979be}, 

\begin{equation}\label{eq:eedecays}
  \Gamma_{ee}(n^3S_1) = \frac{4\alpha^2e_t^2}{M_{nS}^2}|R_{nS}(0)|^2\left(1-\frac{16}{3}\frac{\alpha_s}{\pi}\right)
\end{equation}

\begin{equation}\label{eq:eedecayp}
  \Gamma_{ee}(n^3D_1) = \frac{25\alpha^2e_t^2}{2M_{nD}^2m_t^4}|R_{nD}^{\prime\prime}(0)|^2\left(1-\frac{16}{3}\frac{\alpha_s}{\pi}\right)
\end{equation}

Therefore, we obtain the numerical results,
\begin{eqnarray}
  \Gamma_{ee}(1^3S_1) &=& 39.89~\textrm{keV} \nonumber\\
  \Gamma_{ee}(2^3S_1) &=& 5.19~\textrm{keV} \nonumber\\
  \Gamma_{ee}(3^3S_1) &=& 1.82~\textrm{keV} \nonumber\\
  \Gamma_{ee}(1^3D_1) &=& 6.0\times10^{-6}~\textrm{keV} \\
\end{eqnarray}

\subsection{Muti-photon decays}

We can find the analytical formulas for the muti-photon decay widths in the Refs.
\cite{Kwong:1988ae,Kwong:1987ak},

\begin{eqnarray}\label{eq:twophotons}
  &\Gamma(n^1S_0\rightarrow\gamma\gamma)=\frac{3\alpha^2e_t^4}{m_t^2}|R_{nS}(0)|^2
  \left[1+\frac{\alpha_s}{\pi}\left(\frac{\pi^2}{3}-\frac{20}{3}\right)\right] \nonumber\\
  &\Gamma(n^3P_0\rightarrow\gamma\gamma)=\frac{27\alpha^2e_t^4}{m_t^4}|R_{nP}^\prime(0)|^2
  \left[1+\frac{\alpha_s}{\pi}\left(\frac{\pi^2}{3}-\frac{28}{9}\right)\right] \nonumber\\
  &\Gamma(n^3P_2\rightarrow\gamma\gamma)=\frac{36\alpha^2e_t^4}{5m_t^4}|R_{nP}^\prime(0)|^2
  \left[1-\frac{\alpha_s}{\pi}\frac{16}{3}\right] \nonumber\\
  &\Gamma(n^3S_1\rightarrow\gamma\gamma\gamma)=\frac{4(\pi^2-9)\alpha^3e_t^6}{3\pi m_t^2}|R_{nS}(0)|^2 \left[1-\frac{12.6\alpha_s}{\pi}\right]
\end{eqnarray}
In the nonrelativistic limit, $m_t$ can be replaced by
$M/2$, where M is the mass of the corresponding toponium state.
The results are written as below,
\begin{eqnarray}
  \Gamma(1^1S_0\rightarrow\gamma\gamma) &=& 57.18~\textrm{keV} \nonumber\\
  \Gamma(2^1S_0\rightarrow\gamma\gamma) &=& 7.43~\textrm{keV} \nonumber\\
  \Gamma(1^3P_0\rightarrow\gamma\gamma) &=& 0.13~\textrm{keV} \nonumber\\
  \Gamma(1^3P_2\rightarrow\gamma\gamma) &=& 29~\textrm{eV} \nonumber\\
  \Gamma(1^3S_1\rightarrow\gamma\gamma\gamma) &=& 15~\textrm{eV} \nonumber\\
  \Gamma(2^3S_1\rightarrow\gamma\gamma\gamma) &=& 2~\textrm{eV} \\
\end{eqnarray}

\subsection{Gluonic decays}

Owing to the duality, the total widths of hadronic decays are determined by the gluonic decays. The formulas are also accounted from the previous works,
\begin{eqnarray}\label{eq:gluonsdecay}
  &\Gamma(n^1S_0\rightarrow gg) = \frac{2\alpha_s^2}{3m_t^2}|R_{nS}(0)|^2\left(1+\frac{3.99\alpha_s}{\pi}\right)    \nonumber\\
  &\Gamma(n^3P_0\rightarrow gg) =\frac{6\alpha_s^2}{m_t^4}|R_{nP}^\prime(0)|^2  \nonumber\\
  &\Gamma(n^3P_2\rightarrow gg) =\frac{8\alpha_s^2}{5m_t^4}|R_{nP}^\prime(0)|^2  \nonumber\\  
  &\Gamma(n^1D_2\rightarrow gg) =\frac{2\alpha_s^2}{3\pi m_t^6}|R_{nD}^{\prime\prime}(0)|^2  \nonumber\\
  &\Gamma(n^3S_1\rightarrow ggg) = \frac{10(\pi^2-9)\alpha_s^3}{81\pi m_t^2}|R_{nS}(0)|^2\left(1-\frac{6.06\alpha_s}{\pi}\right)    \nonumber\\
\end{eqnarray}

The results are
listed in Table.\ref{tab:gluonicdec}.

{
\begin{table}[!hbp]
 \begin{center}
\caption{\label{tab:gluonicdec} Gluonic decay widths 
with QCD corrections for toponium states in units of $\textrm{keV}$. In each cell, the number outside the bracket is for the leading order, and inside for the next-to-leading order.}
\begin{tabular}{c|ccccc|cc}
\hline
\hline
states&$1^1S_0$&$2^1S_0$&$1^3P_0$&$1^3P_2$&$1^1D_2$&$1^3S_1$&$2^3S_1$ \\
$\Gamma$&$12960.8(14574.0)$&$1685.1(1894.8)$&$26.4$&$7.0$&$0.2~\textrm{eV}$&$65.0(52.7)$&$8.5(6.9)$ \\
\hline
\end{tabular}
\end{center}
\end{table}}

\section{Hadronic production}

The  dominant subprocess 
is the gluon fusion for the $\eta_t(1^1S_0)$ production in high–energy proton-proton collisions. The lowest order cross section can be expressed as follows which is in direct proportion to the gluonic width of $\eta_t$ \cite{Spira:1995rr,Zhang:2016xei}.

\begin{equation}\label{eq:hadoniccs}
  \sigma^{0}(pp\rightarrow 1^1S_0 
  ) = \frac{\Gamma_{LO}(1^1S_0 \rightarrow gg)}{M s}\frac{\pi^2}{8}\int_{M^2/s}^{1}\frac{dx}{x}g(x)g(\frac{M^2}{sx}),
\end{equation}
where NNPDF4.0 set of parton distributions (PDFs) at LO has been adopt through the ManeParse program, which is a Mathematica package to  provide access to PDFs \cite{Clark:2016jgm}.


{
\begin{table}[!hbp]
 \begin{center}
\caption{\label{tab:hadpro} Theoretical cross sections (in units of $\textrm{pb}$) at the $pp$ colliders for $1^1S_0$ toponium states.}
\begin{tabular}{c|cccccc}
\hline
\hline
$\sqrt{s}(\textrm{TeV})$&$13$&$13.6$&$14$&$27$&$50$&$100$ \\
$\sigma^{0}(\textrm{pb})$&$6.1$&$6.7$&$7.2$&$27.2$&$80.9$&$239.7$ \\
\hline
\end{tabular}
\end{center}
\end{table}}

The results
listed in Table.\ref{tab:hadpro}.
The value of the cross section at $\sqrt{s}=13~\textrm{TeV}$ is consistent with the prediction in Ref.\cite{Fuks:2021xje} and compile with the data at the LHC\cite{CMS:2024ynj}.

\section{Summary}\label{sec:sum}

Predictions of the mass spectrum, especially the ground state masses, are significantly affected by the scale choice \cite{Llanes-Estrada:2024phk}. In future, the parameters in the potential would be well fit by more precision data for toponium mass. As the simplest two-body system of quantum chromodynamics, a series of subsequent theoretical and experimental comparisons of decay, production help to understand the relevant physics of heavy quarkonium.

\begin{acknowledgments}
This work was supported by the National Natural Science Foundation of China (No 11705078).

\end{acknowledgments}

\clearpage
\providecommand{\href}[2]{#2}\begingroup\raggedright\endgroup

\end{document}